\documentclass[article]{revtex4}
\usepackage{graphicx}
\usepackage{amssymb}
\usepackage{amsmath}
\usepackage{bm}
\usepackage{hyperref}
\usepackage{epstopdf}
\usepackage[dvips]{color}

\begin{document}

\title{Viscous dark matter growth in (neo-)Newtonian cosmology }

\author{H. Velten$^{1,2}$\email{velten@physik.uni-bielefeld.br}, D. J. Schwarz$^1$\email{dschwarz@physik.uni-bielefeld.de}, J. C. Fabris$^{2}$\email{fabris@pq.cnpq.br} and W. Zimdahl$^{2}$\email{winfried.zimdahl@pq.cnpq.br}}

\affiliation{$^1$Fakult\"at f\"ur Physik, Universit\"at Bielefeld, Postfach 100131, 33501 Bielefeld, Germany\\$^2$Departamento de F\'sica, Universidade Federal do Esp\'irito Santo,
Vit\'oria, 29060-900, Esp\'irito Santo, Brazil}

\date{\today}

\begin{abstract}
We assume cold dark matter to possess a small bulk-viscous pressure which typically attenuates the
growth of inhomogeneities. Explicit calculations, based on Eckart's theory of dissipative processes,
reveal that for viscous cold dark matter the usual Newtonian approximation for perturbation scales
smaller than the Hubble scale is no longer valid. We advocate the use of a neo-Newtonian
approach which consistently incorporates pressure effects into the fluid dynamics and
correctly reproduces the general relativistic dynamics. This result is of interest for
numerical simulations of nonlinear structure formation involving nonstandard
dark-matter fluids.
We obtain upper limits on the magnitude of the viscous pressure by requiring that relevant perturbation amplitudes
should grow sufficiently to enter the nonlinear stage.

\end{abstract}

\maketitle

\section{Introduction}

The standard scenario of cosmic structure formation relies on the $\Lambda$CDM model.
Although this model fits most of the data surprisingly well, there remain, e.g., the puzzle of
missing satellites \cite{Bullock} and the cusp-core problem \cite{cusp}.
More recently, the Planck satellite has observed much fewer clusters than predicted \cite{Planck2013SZ}. This would be a ``cluster version" of the
missing satellites problem. All these apparent shortcomings of the $\Lambda$CDM model have in common
that pressureless cold dark matter (CDM) leads to an excess of structure and clustering.
Now, although dark matter is certainly very close to be pressureless, its equation of state (EoS) parameter $w_{\rm dm}$ is not necessarily zero exactly \cite{DMEoS}. Indeed, the EoS parameter of a
nonrelativistic gas in thermal equilibrium calculated from kinetic theory is of the order of
$10^{-6}$ \cite{kinetictheory}.
There are also investigations about effects of a nonvanishing EoS parameter using observations of gravitational lensing \cite{faber}, gravitational dynamics at galactic scales \cite{Bharadwaj}, at galaxy cluster scales \cite{mariano} and at cosmological scales \cite{EoSDMnegative}.

Structure formation on scales smaller than the Hubble scale is believed
to be describable within a Newtonian approximation.
Newtonian cosmology is well established since many decades \cite{Milne}.
However, its applicability is related to the circumstance that CDM is assumed to be pressureless.
The results of N-body simulations which are crucial for modern cosmology are based on
pure Newtonian codes.
But as soon as there appears a nonvanishing dark-matter pressure contribution, even if it is a tiny one, the applicability of the Newtonian approximation has to be reconsidered.

In this paper we address the question to what extent a Newtonian approximation remains reliable
if (small) deviations from $w_{\rm dm} = 0$ are taken into account.
Whether or not Newtonian simulations
are missing any (relativistic) aspect at the largest scales is a matter of intense debate
in the literature \cite{NewSimuTheory}.
Our focus here will be on scales smaller than the Hubble scale. In particular, we aim at clarifying
the potential role of small pressure effects for the evolution of dark matter inhomogeneities.
We shall demonstrate that in this situation one can no longer trust the standard Newtonian approximation.  On the other hand, a neo-Newtonian generalization which is able to incorporate pressure effects, yields results that well approximate those of an exact general relativistic treatment.

Following a previous analysis \cite{dominikVelten2012}, we shall consider viscous CDM (vCDM) in which dissipative effects give rise to a small bulk viscous pressure.
While physically an effective bulk viscous pressure should be the result of a
self-interaction within the dark matter, our approach will be entirely
phenomenological, using  Eckart's theory \cite{eckart} of dissipative processes.
We do not assume that this viscous pressure is
responsible for the accelerated expansion of the Universe as in \cite{viscous1, winfried, dominik}.
Instead, it is supposed to describe a small deviation from the $\Lambda$CDM model which is called
$\Lambda$vCDM model.

The structure of the paper is as follows.
In Sec.~\ref{newtoncos} we review the Newtonian and the neo-Newtonian approaches to cosmology.
Section \ref{background} is devoted to the introduction of the
viscous-matter cosmology.
Perturbation schemes for the Newtonian, neo-Newtonian and general relativistic levels of description are developed in Sec.~\ref{perturbations}.
The results for the growth of linear viscous dark-matter perturbations are presented in Sec.~\ref{growth}. A summary of the paper is given in Sec.~\ref{discussion}.

\section{Newtonian and neo-Newtonian Cosmologies}
\label{newtoncos}

The Newtonian cosmology has been established in the 1930s by Milne and McCrea \cite{Milne}. It is described by the equations
\begin{equation}\label{continuity1}
\dot{\rho} + \nabla_r \cdot \left( \rho {\bf v} \right)=0\,,
\end{equation}

\begin{equation}\label{euler1}
\dot{ {\bf v}}+ ({\bf v} \cdot \nabla_r) {\bf v} = - \nabla_r \Psi - \frac{ \nabla_r P}{\rho},
\end{equation}
and
\begin{equation}
\label{poisson1}\nabla^{2}_r \Psi = 4 \pi G \rho,
\end{equation}
where $\rho$ is the energy density, $\mathbf{v}$ is the fluid velocity, $P$ is the pressure and $\Psi$ is the gravitational potential. For our expanding Universe the background velocity field of the matter particles obeys the Hubble law ${\bf v}=H(t) {\bf r}(t)$. Here, $\mathbf{r}(t) = \mathbf{r}_{0}\frac{a(t)}{a_{0}}$ and $H = \frac{\dot{a}}{a}$. A dot on top of a variable denotes a partial derivative with respect to the cosmic time $t$.
The subscript $r$ denotes Eulerian coordinates. Equations (\ref{continuity1}) and (\ref{euler1}) are the continuity and the Euler equations, respectively. They provide a fluid picture of the cosmic medium which is gravitationally self-interacting via the Poisson equation (\ref{poisson1}).
In the Newtonian cosmology the Friedmann equations read

\begin{equation}
\frac{\dot{a}^2}{a^2}+\frac{\left(-2E\right)}{a^2}=\frac{8\pi G}{3} \rho \hspace{1cm}{\rm and}\hspace{1cm}\dot{H}+H^2=-\frac{4\pi G}{3}\rho,
\end{equation}
where $E$ is a constant of integration which plays the role of the energy of the expanding system. In this Newtonian treatment the pressure is not dynamically relevant for the homogeneous and isotropic background. With Newtonian cosmology it is not possible to model a radiation dominated phase or even to
study a late time dark energy dominated epoch. This approach is restricted to a description of an
Einstein-de Sitter universe which is a model dominated by pressureless matter.

A simple way to include the effects of the pressure and, at the same time, keep the simplicity of the Newtonian
physics is to use the neo-Newtonian (or pseudo-Newtonian) equations developed during the 1950s by McCrea
\cite{McCrea1951} and by Harrison in the 1960s \cite{Harrison1965}. Later, during the 1990s, an important
analysis concerning the perturbative behavior of Harrison's neo-Newtonian equations helped to set the final form
for the system of equations in this approach \cite{Lima1997} (see also \cite{RRRR2003}). This set of equations
reads
\begin{equation}\label{continuity2}
\dot{ \rho}+ \nabla_r \cdot \left( \rho {\bf v} \right)+P\nabla_r \cdot {\bf v}=0\,,
\end{equation}
\begin{equation}\label{euler2}
\dot{ {\bf v}} + ({\bf v} \cdot  \nabla_r ){\bf v} = - \nabla_r \Psi   - \frac{ \nabla_r  P}{\rho + P}\,,
\end{equation}
\begin{equation}
\label{poisson2}\nabla^{2}_r \Psi = 4 \pi G \left[ \rho + 3 P\right].
\end{equation}

Combining eqs.~(\ref{continuity2}), (\ref{euler2}) and (\ref{poisson2}) one obtains equations for the expansion of the homogeneous and isotropic background that are exactly the same as the relativistic Friedmann equations

\begin{equation}
\frac{\dot{a}^2}{a^2}+\frac{\left(-2E\right)}{a^2}=\frac{8\pi G}{3} \rho \hspace{1cm}{\rm and}\hspace{1cm}\dot{H}+H^2=-\frac{4\pi G}{3}(\rho+3P).
\end{equation}

\section{Background evolution in the case of viscous cold dark matter}

\label{background}

Standard cosmology is based on the notion of perfect fluids and the effective pressure of the fluid is interpreted as its equilibrium (kinetic) pressure. However, since the Universe is made of real components rather than ideal ones, one can go a step further including dissipative mechanisms in the above approach. In a homogeneous and isotropic background, i.e.~if the cosmological principle holds, directional dissipative processes (shear and heat conduction) cannot play a decisive role in the cosmic dynamics. On the other hand, the expansion can induce a small deviation from equilibrium in the form of a bulk viscosity.

Our cosmological model is very similar to the standard $\Lambda$CDM scenario.
We assume a flat background expansion
of a universe consisting of baryons, radiation, a cosmological constant and viscous Cold Dark Matter (vCDM). Attributing a viscosity to cold dark matter represents the only modification compared with the $\Lambda$CDM model. As in the standard case, the components are assumed to obey separately the energy conservation laws
\begin{equation} \label{Cont}
\dot{\rho}_{A}+3H(\rho_{A}+P_{A})=0,\qquad (A = {\rm b, r, v}, \Lambda)\,,
\end{equation}
where the subscripts ${\rm b}$, ${\rm r}$, ${\rm v} $ and $\Lambda$ denote baryons, radiation, viscous dark matter and cosmological constant, respectively.
The total energy is $\rho = \rho_{\rm b}+ \rho_{\rm r} + \rho_{\rm v} + \rho_{\Lambda}$ and the total pressure
$P = P_{\rm b}+ P_{\rm r} + P_{\rm v} + P_{\Lambda} = \frac{1}{3}\rho_{\rm r} + P_{\rm v} - \rho_{\Lambda}$.
Note that different from the Newtonian description based on (\ref{continuity1}), (\ref{euler1}) and (\ref{poisson1}),
the neo-Newtonian approach incorporates the pressure contributions of the different components already at
the background level.
It would not have been possible, e.g., to include a radiation component with $P_{\rm r}=\rho_{\rm r}/3$ into the traditional Newtonian description.

The background expansion of our model is given by
\begin{equation}
H^{2}=H^{2}_{0}\left[\Omega_{\rm b0}(1+z)^3+\Omega_{\rm r0}(1+z)^4+\Omega_{\rm v}(z)+\Omega_{\Lambda}\right],
\label{H}
\end{equation}
where
\begin{equation}\label{}
\Omega_{\rm b0} = \frac{8\pi G \rho_{\rm b0}}{3 H_{0}^{2}}\,,\quad \Omega_{\rm r0} = \frac{8\pi G \rho_{\rm r0}}{3 H_{0}^{2}}\,,\quad \Omega_{\rm v} = \frac{8\pi G \rho_{\rm v}}{3 H^{2}_0}\,,\quad \Omega_{\Lambda} = \frac{8\pi G \rho_{\Lambda}}{3 H^{2}_0}\,.
\end{equation}
For $H_0, \Omega_{b0}$ and $\Omega_{r0}$
we will assume the values reported by the Planck satellite mission \cite{Planck2013CP}.

In order to obtain the function $\Omega_{\rm v}\left(z\right)$ we have to specify the pressure $P_{\rm v}$ of the vCDM and solve its continuity equation. We emphasize again that in the traditional Newtonian cosmology
(superscript $N$) the pressure is not relevant for the homogeneous and isotropic background and we have $\Omega_{\rm{v}}^{N}(z)=\Omega_{\rm{v}0}(1+z)^3$. However, the neo-Newtonian and the relativistic theories take into account the dynamical effects of the pressure for the background and the evolution of the viscous dark-matter density will obey the conservation equation (\ref{Cont}).

To study bulk viscous phenomena we decompose the pressure $P_{\rm v}$ according to
\begin{equation}\label{effectivePressure}
P_{\rm v}=p_{\rm k} + \Pi,
\end{equation}
where $p_{\rm k}\equiv p_{\rm k}(\rho)$ is the kinetic (adiabatic) pressure and $\Pi$ is the bulk viscous (non-adiabatic) one.

The form of $\Pi$ is determined by Eckart's theory of dissipative processes \cite{eckart}. Our goal in this work is to understand what is the appropriate Newtonian approximation for the study of bulk viscous effects until the onset of non-linear structure formation.

For simplicity we assume that the viscous dark-matter fluid has a vanishing kinetic pressure $p_{\rm k}=0$.
Then its only pressure contribution is a (small) viscous pressure $\Pi\neq 0$.
One reason for this assumption is that the effects of a small adiabatic pressure give rise to the famous
Jeans instability criterion, i.e. the formation of structure is prevented below the so-called Jeans scale.
In contrast to the adiabatic pressure, the effect of bulk pressure can damp density fluctuations.
We recall that we regard the bulk viscous pressure as an effective description of deviations from the standard
pressureless CDM fluid. Note that in kinetic gas theory with standard interactions a viscous pressure is a correction to the adiabatic pressure.

In the present situation, the conservation law (\ref{Cont}) in terms of dimensionless quantities reads
\begin{equation}\label{backgroundviscous}
-(1+z)\frac{d\Omega_{\rm v}(z)}{dz}+3\Omega_{\rm v}(z)+\frac{\tilde{\Pi}}{3}=0\,, \qquad \tilde{\Pi} =\frac{24\pi G \Pi}{H^2_0}.
\end{equation}
These relations are valid for any pressure $\Pi$. In Eckart's theory one has
\begin{equation}
\Pi=-\xi u^{\gamma}_{; \gamma}\,,
\end{equation}
where $\xi$ is the coefficient of bulk viscosity and $u^{\gamma}$ is the fluid $4-$velocity (greek indices run over 0,1,2,3). In a homogeneous and isotropic universe the expansion scalar becomes $u^{\gamma}_{; \gamma}=3H$.
The corresponding relation in a Newtonian context is $v^{a}_{,a} = 3 H$ (latin indices run over $1, 2, 3$).

In the absence of a microscopic theory for the cosmological bulk viscosity, a standard phenomenological choice for $\xi$ is the energy density dependence
\begin{equation}
\xi=\xi_0\left(\frac{\rho_{\rm v}}{\rho_{\rm  v0}}\right)^{\nu},
\label{xi}
\end{equation}
with $\xi_0$ and $\nu$ being constants and $\rho_{\rm v0}$ being the density of the viscous fluid today. This means that the current viscosity of the dark matter fluid is given by the parameter $\xi_0$. This density dependence of $\xi$ is motivated by the fact that transport coefficients derived in kinetic theory depend on powers of the temperature of the fluid. One can find in the literature approaches where $\xi\equiv\xi(H, H^2)$ implying that the viscosity depends indirectly on the density of the other components. In this case, it is mandatory to assume a coupling between the components violating the assumption of separated energy conservation.

Assuming (\ref{xi}), the energy-conservation equation for the viscous dark matter, in terms of the redshift $z$, is written as
\begin{equation}
(1+z)\frac{d \Omega_{\rm v}(z)}{dz}-3\Omega_{\rm v}\left(z\right)+\tilde{\xi}\left(\frac{\Omega_{\rm v}\left(z\right)}{\Omega_{\rm v0}}\right)^{\nu}\left[\Omega_{\rm r0}(1+z)^{4}+\Omega_{\rm b0}(1+z)^{3}+\Omega_{\rm v}\left(z\right)+\Omega_{\Lambda}\right]^{1/2}=0,
\label{model2}
\end{equation}
where
\begin{equation}
\tilde{\xi} = \frac{24\pi G \xi_0}{H_0}.
\end{equation}
As initial condition we set $\Omega_{\rm v}(z=0)=\Omega_{{\rm v}0} = \Omega_{\rm m}=0.12029 h^{-2}$ \cite{Planck2013CP}.
Below, we will show results for the viscosity of dark matter in terms of the parameter $\tilde{\xi}$ and for this reason it is important to relate this quantity to the vCDM equation of state parameter today via
\begin{equation} \label{wv0}
w_{\rm v 0}=-\frac{\tilde{\xi}}{3 \Omega_{\rm v 0}}.
\end{equation}
Note that CDM is fully recovered if $\tilde{\xi}=0$.

\section{Perturbative dynamics of viscous cosmic fluids}
\label{perturbations}

Cosmic structures are formed under gravitational agglomeration of dark matter halos and the subsequent accretion of baryons during the matter dominated epoch. The galaxy clustering patterns, observed by large-scale surveys, can be directly compared to theoretical predictions by using either N-body or hydrodynamical simulations. The latter uses fluid dynamics, i.e. continuity and Euler equations coupled to the Poisson equation.

During the radiation dominated epoch the Hubble drag stagnates such structures. When the background becomes matter dominated (for the concordance model this occurs at $z\sim3000$), the fractional density perturbations $\Delta$, which subsequently evolve into the standard CDM halos, start to grow linearly with the scale factor, $\Delta \sim a$. The smallest scales become nonlinear first, decoupling from the Hubble flow. The perturbations on the remaining sub-horizon scales which are still in the linear regime continue to grow until dark energy accelerates the background expansion at late times causing a a suppression of the growth of $\Delta$. For pressureless CDM this picture is achieved in both Newtonian and relativistic formulations but for the case of vCDM we could expect that the dissipation leads to an extra suppression mechanism.

In the next subsections we derive the equations needed in order to study the evolution of sub-horizon viscous dark matter halos. We calculate the evolution of the density contrast within the Newtonian, neo-Newtonian and general relativistic cases.

\subsection{Viscous fluid in Newtonian description}

In order to understand the structure formation it is necessary to apply cosmological perturbation theory. We briefly describe the procedure. Firstly, we decompose eqs.~(\ref{continuity1}), (\ref{euler1}) and (\ref{poisson1}) into homogeneous and isotropic background quantities and first order perturbations. We write the quantities $f = \left\{ \rho, {\bf v}, p_{\rm k}, \xi, \Psi \right\}$ as $f=f+\delta f$, where $f$ denotes the background value and $\delta f$ means a small fluctuation. At linear order $\delta{f}/f \ll 1$. We also Fourier transform the perturbations to the wave-number ($k$) space. It is also necessary to replace the Euler coordinates by the Lagrangian ones (in Lagrangian coordinates ${\bf q}={\bf r}/a$). Note that this treatment is valid for any effective pressure $P$.

For the pressure $P$ in (\ref{euler1}) we have
\begin{equation}\label{Eckart}
P=p_k+\Pi=p_k -\xi \theta =- \xi \nabla_r \cdot {\bf v},
\end{equation}
where  $\theta$ represents the system's volume expansion. It is clear that under this redefinition Eq. (\ref{euler1}) becomes the standard Navier-Stokes equation (here without shear viscosity)
\begin{equation}\label{NS1}
\dot{ {\bf v}}+ ({\bf v} \cdot \nabla_r ){\bf v} = - \nabla_r \Psi - \frac{\nabla_r p_{\rm k} }{\rho} +
\frac{ \nabla_r \left(\xi\nabla_r \cdot {\bf v}\right)}{\rho},
\end{equation}
which is a generalization of the Euler equation.

In the expanding, homogeneous and isotropic background one has $\nabla_r \cdot {\bf v}=3H(t)$.
Note that $\xi$ appears inside the gradient operator in equation (\ref{NS1}). Since the background
viscosity is only time-dependent ($\xi\equiv \xi(t)$) one often writes $\xi  \nabla_r \left(\nabla_r \cdot
{\bf v}\right)/\rho$ instead of the last term of (\ref{NS1}). However, at linear order, perturbations of
$\xi$ generally depend on the position, i.e., $\delta{\xi}\equiv\delta\xi({\bf r}, t)$.

Perturbing the pressure $P$ given by (\ref{Eckart}), we find
\begin{equation}
\delta P = \left(\frac{\partial p_{\rm k}}{\partial \rho}\right) \delta \rho+\delta \Pi=c_s^2 \rho \Delta+\nu w_{\rm v}\rho \Delta-\frac{w_{\rm v}}{3H}\rho \dot{\Delta}.
\end{equation}
We have used the first-order Eq. (\ref{continuity1}) in order to eliminate the term
$\nabla_r \cdot (\delta {\bf v})$ (present) in $\delta\Pi$.

The contribution of the kinetic pressure to the perturbative dynamics occurs via the definition of the adiabatic sound speed $c_s^2 =\delta p_{\rm k}/ \delta \rho$. We have also defined the viscous equation of state parameter
\begin{equation}
 w_{\rm v}=\frac{\Pi}{\rho}=-\frac{3H\xi}{\rho} .
\end{equation}
A standard result for the perturbations in the Newtonian theory is the equation for the density contrast $\Delta = \delta{\rho}/\rho$
\begin{equation}\label{eqDelta1}
\ddot{\Delta}+2H\dot{\Delta}-4\pi G \rho\Delta=-\frac{k^2}{a^2}\frac{\delta P}{\rho},
\end{equation}
which is valid for any effective pressure $P$. Hence, combining the above relations we find
\begin{equation}\label{eqDeltaViscous1}
\ddot{\Delta}+\left(2H-\frac{w_{\rm v} k^2}{3 H a^2}\right)\dot{\Delta}+\left(\frac{c_s^2 k^2}{a^2}+\nu\frac{ w_{\rm v}k^2}{a^2}-4\pi G\rho\right)\Delta=0.
\end{equation}
Equation (\ref{eqDeltaViscous1}) differs from Eq. (36) in \cite{carlevaro} because we took into account perturbations of the bulk viscous coefficient $\delta\xi = \nu \xi \Delta$ which generates the term proportional to $\nu$. This is the first new aspect of this paper.

In the following we work with the scale factor $a$ as our dynamical variable rather than the cosmic time $t$. Hence, Eq. (\ref{eqDeltaViscous1}) reads
\begin{equation}\label{eqDeltaViscous1a}
a^2 \Delta^{\prime\prime}+\left(\frac{a H^{\prime}}{H}+3-\frac{w_{\rm v} k^2}{3 H^2 a^2}\right)a\Delta^{\prime}+\left(\frac{c_s^2 k^2}{H^2a^2}+\nu\frac{ w_{\rm v}k^2}{H^2a^2}-\frac{3}{2}\frac{H_0^2 \Omega}{H^2}\right)\Delta=0,
\end{equation}
where the prime denotes a derivative with respect to the scale factor. For the
${\rm v}$CDM fluid we identify $\Omega=\Omega_{\rm v0}$ and $c^2_s=0$ in the above equation.

\subsection{Viscous fluids in neo-Newtonian cosmology}

The goal of this section is to develop the Newtonian perturbative dynamics including the effects of the pressure properly. Of course, this is unnecessary for the standard CDM fluid ($P_{\rm{CDM}}=0$), but it is obvious that for our viscous fluid $P_{\rm{v}}\neq 0$. We obtain a neo-Newtonian perturbative dynamics for a general fluid with additional viscous pressure $\Pi$ as in (\ref{effectivePressure}). The perturbed equations (\ref{continuity2}) --
(\ref{poisson2}) read, respectively
\begin{equation}\label{continuityp}
\dot{\Delta}  - 3Hw \Delta + (1+w)\frac{\nabla \cdot \delta{\bf v}}{a}+\frac{3 H}{\rho}\delta P=0,
\end{equation}
\begin{equation}\label{Eulerp}
\dot{ \delta{\bf v}} + H \delta{\bf v}= -\frac{\nabla \delta\Psi}{a} -\frac{\nabla \delta P}{a\rho(1+w)},
\end{equation}
\begin{equation}\label{Poissonp}
\nabla^2 \,\delta\Psi = 4\pi G a^2 \rho\Delta+12\pi G a^2\,\delta P.
\end{equation}
Note that the above set of equation is valid for any pressure $P$ with equation of state parameter $w=P/\rho$. Combining equations (\ref{continuityp})-(\ref{Poissonp}), we find the following evolution for the density contrast
\begin{eqnarray}
\ddot{\Delta}+\left(2H-3 H w-\frac{\dot{w}}{1+w}\right)\dot{\Delta}+\left[-4\pi G \rho \left(1+w\right)-6 H^2 w-3\dot{H}w-\frac{3H \dot{w}}{1+w}\right]\Delta = \nonumber\\
\frac{\nabla^2_q \delta P}{a^2\rho} -3H\frac{\dot{\delta P}}{\rho}+\frac{\delta P}{\rho}\left[12\pi G \rho(1+w)-9H^2 w+\frac{3H\dot{w}}{1+w}-15 H^2-3\dot{H}\right].
\end{eqnarray}
Again, this equation is valid for any pressure $P$ with perturbation $\delta P=c^2_s \delta \rho +\delta\Pi$.
The quantities on the right-hand side can be written in cosmological units and the dynamical variable changed to the scale factor, resulting in
\begin{eqnarray} \label{Deltatimea}
a^2\Delta^{\prime\prime}+\left(3+\frac{a H^{\prime}}{H} -3w-\frac{a{w}^{\prime}}{1+w}\right)a\Delta^{\prime}+\left[-\frac{3 H^2_0 \Omega}{2H^2} \left(1+w \right)-6 w-\frac{3H^{\prime} a w}{H}-\frac{3 a w^{\prime}}{1+w}\right]\Delta = \nonumber \\
-k^2\frac{\tilde{\delta P}}{9 a^2H^2\Omega} -a\frac{\tilde{\delta P}^{\prime}}{3\Omega}+\frac{\tilde{\delta P}}{9\Omega}\left[\frac{9 H^2_0 \Omega}{2H^2}(1+w)-9w-\frac{3aw^{\prime}}{(1+w)}-15-\frac{3aH^{\prime}}{H}\right] .
\end{eqnarray}
Here, we wrote the pressure in terms of the dimensionless quantity $\tilde{\delta P}=24 \pi G \delta P / H^2_0$. We now need the proper form of the perturbation $\tilde{\delta P}$ in equation (\ref{Deltatimea}). For typical matter components, e.g. baryons and CDM, the kinetic pressure is set to be zero ($p_{k}=0$) and therefore we adopt $c_s^2=0$. However, our dark matter fluid still has a nonvanishing viscous pressure $\Pi$, i.e. $w=w_{\rm v}$. Hence, we have $\delta P =\delta \Pi$.

Perturbing the bulk-viscous pressure part in (\ref{Eckart}) we have
\begin{equation}
\delta\Pi = -\delta\xi \nabla_{r}  \cdot {\bf v}-\xi \nabla_r \cdot \delta{\bf v}=-3H\delta\xi-\frac{\xi\nabla  \cdot
\delta{\bf v}}{a},
\end{equation}
which, combined with Eq. (\ref{continuityp}) results in
\begin{equation}\label{deltaPiEckarta}
\frac{\delta P}{\rho}=\frac{\delta\Pi}{\rho}=\frac{\tilde{\delta P}}{9\Omega} =\frac{\tilde{\delta\Pi}}{9\Omega} =\frac{w_{\rm v}}{3}\left[\frac{-a \Delta^{\prime}+3 w_{\rm v} \Delta + 3 \nu (1+w_{\rm v})\Delta }{1+2w_{\rm v}}\right].
\end{equation}
With the above relation we can rewrite Eq. (\ref{Deltatimea}) in order to find a
second order differential equation for the density contrast.

For scales $k^{2}\gg a^{2}H^{2}$, the case of interest here, the terms with a factor
$k^{2}/(a^{2}H^{2})$ are expected to dominate the perturbation dynamics.

If we restrict ourselves to terms linear in $w_{\rm v}$, the scale-dependent term in (\ref{Deltatimea}) becomes
\begin{equation}\label{scaleP}
\frac{k^2}{a^2H^2}\frac{\tilde{\delta P}}{9 \Omega}\approx\frac{k^2}{a^2H^2}\frac{w_{\rm v}}{3}\left[-a \Delta^{\prime}+3\nu \Delta\right].
\end{equation}
In fact, in this approximation the scale-dependent neo-Newtonian perturbation terms
coincide with those of the Newtonian equation (\ref{eqDeltaViscous1a}).

\subsection{Viscous fluids in the relativistic cosmology}

For the sake of comparison we now derive the relativistic version of the Meszaros equation for the bulk viscous fluid. Assuming the conformal Newtonian gauge in the absence of anisotropic stresses
\begin{equation}
ds^2=a(\eta)^2\left[-\left(1+2\phi\right)d\eta^2+\left(1-2\phi\right)\delta_{ij}dx^i dx^j\right],
\end{equation}
we can calculate the perturbed part of the energy-momentum balances. These equations read
\begin{equation}\label{rcont}
\dot{\Delta}= -(1+w)\left(\frac{\theta}{a}-3\dot{\phi}\right)+3Hw\Delta-3H\frac{\delta P}{\rho},
\end{equation}
\begin{equation}\label{reuler}
\dot{\theta}= -H(1-3w)\theta-\frac{\dot{w}}{1+w}\theta+\frac{k^2}{a\left(1+w\right)}\frac{\delta P}{\rho}+\frac{k^2}{a}\phi,
\end{equation}
where $\theta=i k^j v_j$ is the divergence of the perturbed fluid velocity.

In order to find a single equation for the density contrast we still need the Poisson equation
\begin{equation}\label{rpoisson}
\frac{k^2}{a^2}\phi +3H\left(\dot{\phi}+H\phi\right)=-4\pi G \rho \Delta.
\end{equation}

For sub-horizon modes we take the large-k limit of the above equation and neglect $\dot{\phi}$ in (\ref{rcont}) \cite{WimpyHalos}. Hence, the relativistic evolution of the density contrast for a general pressure $P$ is
\begin{eqnarray}\label{rdeltaa}
a^2\Delta^{\prime\prime}+\left(3+\frac{aH^{\prime}}{H}-3w\right)a\Delta^{\prime}+\left[-\frac{3 H^2_0  \Omega}{2 H^2}\left(1+w\right)-6w+9w^2-\frac{3aH^{\prime}w}{H}
-3w^{\prime}a\right]\Delta= \nonumber \\
-\frac{k^2}{H^2a^2}\frac{\delta \tilde{P}}{9\Omega}-a\frac{\delta \tilde{P}^{\prime}}{3\Omega}+\frac{\delta \tilde{P}}{9\Omega}\left(-15-\frac{3aH^{\prime}}{H}\right).
\end{eqnarray}
The above equation is valid for any pressure $P$. Identifying now $P=\Pi$ we have, up to first order,
\begin{equation}\label{tn}
\frac{\delta \Pi}{\rho}=\nu w_{\rm{v}} \Delta-\frac{w_{\rm{v}} \theta}{3 H a}+w_{\rm v} \phi.
\end{equation}

With the help of Eq.~(\ref{rcont}) this expression for $\delta \Pi/\rho$ coincides with (\ref{deltaPiEckarta}).
It follows that the fractional pressure perturbations of the neo-Newtonian theory are the same as
those of the relativistic theory.
Limiting ourselves to terms linear in $w_{\rm v}$, we recover (\ref{scaleP}) (recall that 
$\Omega \equiv \Omega_{\rm v}(z=0)$).
Consequently, in lowest order in $w_{\rm v}$, the scale-dependent contributions of the pressure perturbation coincide for all three approaches which provides us with very similar results for the growth of perturbations. 
However, as (\ref{deltaPiEckarta}) shows (cf. the last term in the brackets on the right-hand side of the last equation in (\ref{deltaPiEckarta})), both in the neo-Newtonian and in the relativistic theories there appears a scale-dependent term
of the order $\nu w_{\rm v}^{2}$ which is absent in the Newtonian framework.
Since $k^{2}/a^{2}H^{2} \gg 1$, this term is suggested to contribute even for $w_{\rm v}\ll 1$.
Therefore one expects similar results for the neo-Newtonian and relativistic theories which do not necessarily coincide with Newtonian cosmology, except for $\nu = 0$. This behavior is indeed confirmed by the numerical calculations in the following section. These calculations also imply that corrections of the order of $w_{\rm v }^2$ in
the ``friction'' term are quantitatively less important.
While the scale-dependent terms of the neo-Newtonian and relativistic theories coincide, the second-order equations (\ref{Deltatimea}) and (\ref{rdeltaa}) are different from each other and different results are expected for values of the order of $w_{\rm v} \sim1$.

\section{Results}
\label{growth}

With the equations derived in the last section we are now able to understand the growth of viscous matter inhomogeneities for the Newtonian (N), neo-Newtonian (nN) and relativistic (R) cases. This analysis represents an extension of the findings in ref.\cite{dominikVelten2012}, where only the full theory has been taken into account. The growth of viscous dark matter halos is sensitive to values of the bulk viscosity as small
as $\tilde{\xi}\lesssim 10^{-11}$. Note that for this range the background dynamics of the $\Lambda${\rm v}CDM is indistinguishable from the standard $\Lambda$CDM case. Thus, in practise, both models share the same expansion.
Hence, when solving the equation for the density contrast $\Delta$, we employ the same initial conditions, i.e. we use the CAMB code \cite{CAMB} to set the initial conditions (the amplitude of the power spectrum) at the matter-radiation equality.

Let us investigate the quantitative dependence of the numerical solutions of equations (\ref{eqDeltaViscous1a}), (\ref{Deltatimea}) and (\ref{rdeltaa}) on the parameters $\tilde{\xi}$ and $\nu$, which determine the equation of state parameter $w_{\rm v}$.

In Fig.~1 we plot $\Delta$ as a function of the scale factor for the Newtonian, neo-Newtonian and relativistic
evolution equations. The scales studied here correspond to dwarf galaxies ($k=1000 h$ Mpc$^{-1}$), seen in
the left panel, and galaxy clusters ($k=0.2 h$ Mpc$^{-1}$), seen in the right panel of this same figure.
The horizontal solid line, $\Delta=1$, denotes the limit of validity of the linear theory applied here.
This is the onset of the non-linear regime. The solid curve corresponds to the growth of cold dark matter
halos ($P=0$) for a standard $\Lambda$CDM background. The dashed lines show the evolution of
viscous dark matter halos for different values of the viscosity parameter $\tilde{\xi}$. The parameters
used in each plot are shown in the labels. In Fig.~1 we fix $\nu=0$ and we find an excellent
agreement between the Newtonian, neo-Newtonian and relativistic solutions. We also see that the existence
of nonlinear viscous structures at dwarf galaxy scale demands values of the viscosity of order
$\tilde{\xi} < 5 \times 10^{-11}$.

 Although the same results for the three theories have been obtained for $\nu=0$, equations 
(\ref{eqDeltaViscous1a}), (\ref{Deltatimea}) and (\ref{rdeltaa}), even in this limit they do not coincide exactly. 
However, note that we have used values of $\tilde{\xi} \lesssim 10^{-6}$ and consequently, 
according to (\ref{wv0}), $w_{\rm v0} \lesssim 10^{-6}$. Since we are probing values 
of $w_{\rm v} \ll 1$ the differences, which are of order $O(w,w^2)$, turn out to be negligible.

We perform a similar analysis in Figs.~2 and 3 where we now fix the values $\nu=-1/2$ and $1/4$, respectively.

For the upper (bottom) panels of Fig.~2 we show the results for the dwarf galaxy (galaxy cluster) scale.
Note the agreement between the neo-Newtonian and the relativistic case, while the Newtonian description
does not show the same suppression of growth. For both scales, left panels use the maximum viscosity that
still forms nonlinear structures for the neo-Newtonian and relativistic cases. The right plots show the
maximum viscosity allowed in the Newtonian framework. However, for the given value of $\tilde{\xi}$,
nonlinear structures would never form in the relativistic theory.

In Fig.~3 we do the same analysis for the choice $\nu=1/4$. It is worth noting that now the Newtonian results are
suppressed in comparison with the neo-Newtonian and the relativistic cases.

We also study in Fig.~4 the limiting case $\nu=1/2$. Larger values for $\nu$ cause an inversion of the
evolution of the density. In this case, the viscous fluid would evolve from a ``dark energy'' phase in the past
to a matter scaling today. Of course, such early time behavior is prohibited for a fluid that has to play the role of
dark matter in the universe.

These results confirm our previous expectations (cf.~the comments following Eq.~(\ref{tn})):
all three approaches give similar results for $\nu = 0$, but for $\nu \neq 0$ only the
neo-Newtonian approach is a good approximation to the full relativistic theory.
This is true for small values of $w_{\rm v}$, the only case of interest here. For values of the order
of $w_{\rm v}\sim 1$ the neo-Newtonian theory does no longer approximate the full theory.
In the formal limit $k\rightarrow 0$ (which, of course, is beyond the applicability of the Newtonian theory) and for $w_{\rm v}\ll 1$ we find a similar behavior of cosmological perturbations for all three cases again.
This confirms the dominating role of the $k^{2}/(aH)^{2}$ terms for the dynamics of fluctuations 
at the smallest scales.

\begin{figure}
\begin{center}
\includegraphics[width=0.4\textwidth]{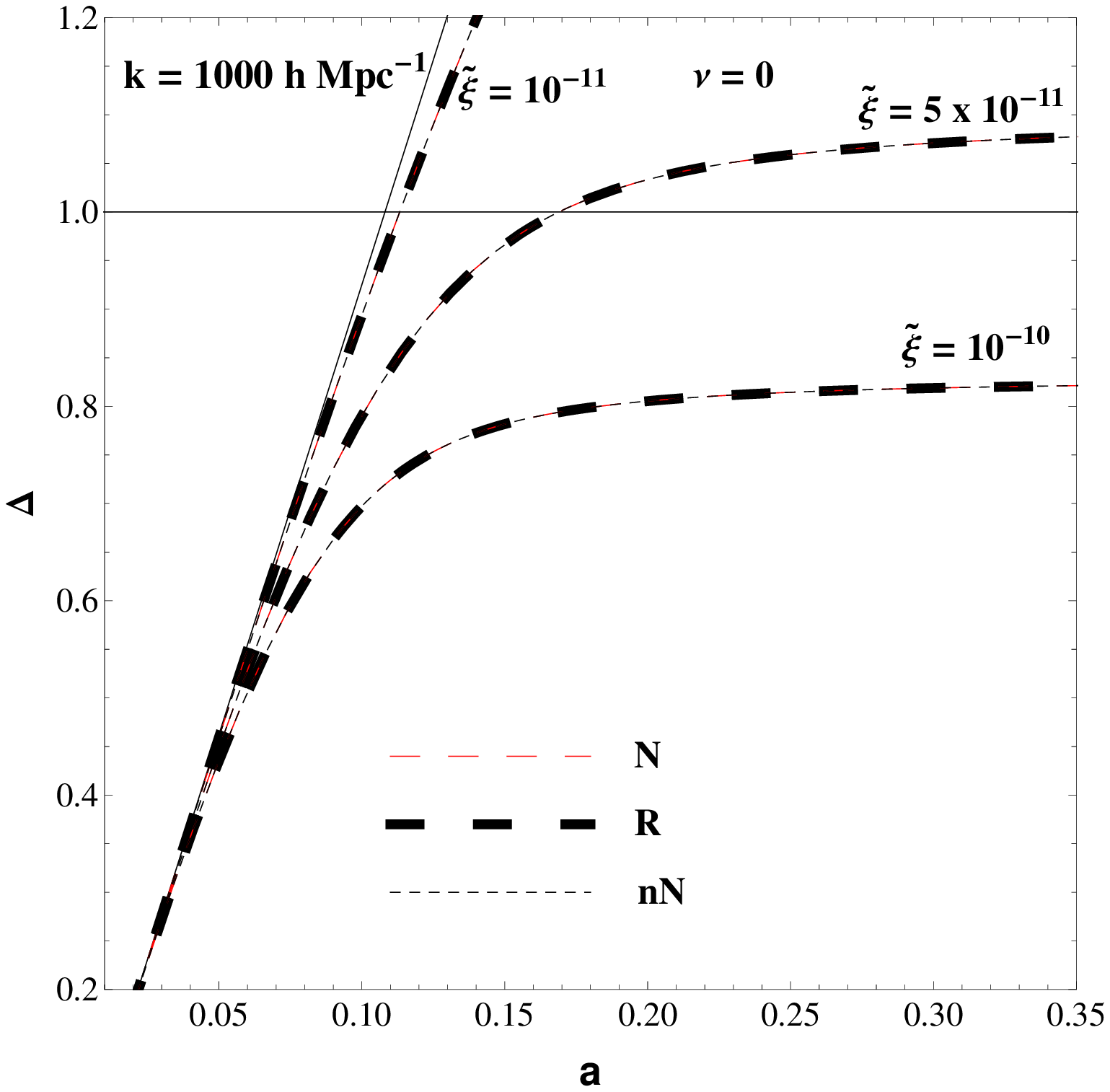}
\includegraphics[width=0.4\textwidth]{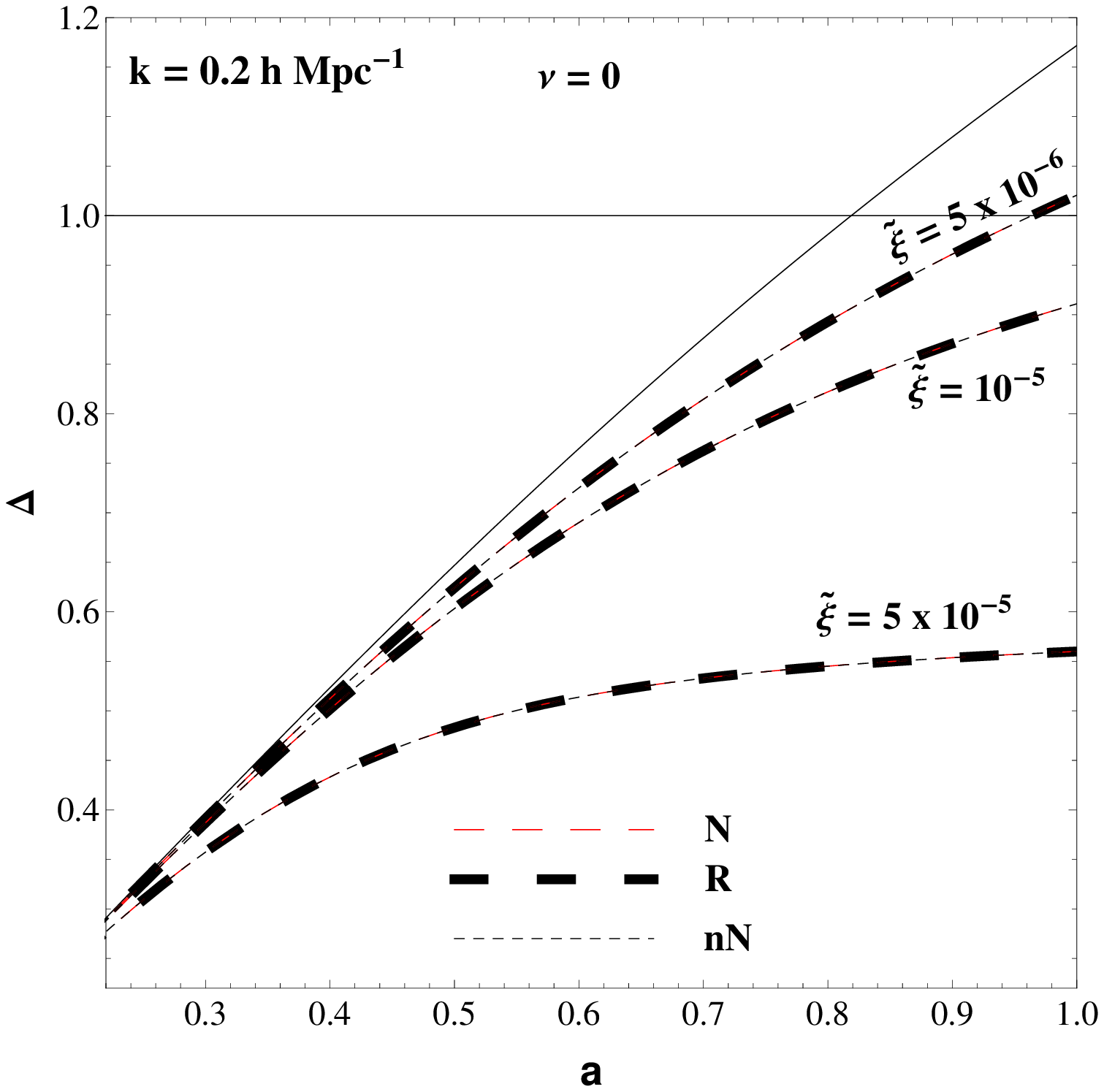}
\caption{Growth of sub-horizon density perturbations for the dwarf galaxy scale $k=1000\, h\,  {\rm Mpc}^{-1}$
(left panel) and the galaxy cluster scale $k=0.2 \, h \, {\rm Mpc}^{-1}$ (right panel) assuming $\nu=0$. The solid line corresponds to the standard $\Lambda$CDM model. The horizontal line indicates
the onset of the nonlinear regime ($\Delta =1$). The Newtonian, neo-Newtonian and relativistic approaches
agree with each other.}
\end{center}
\label{fig1}
\end{figure}

\begin{figure}
\begin{center}
\includegraphics[width=0.35\textwidth]{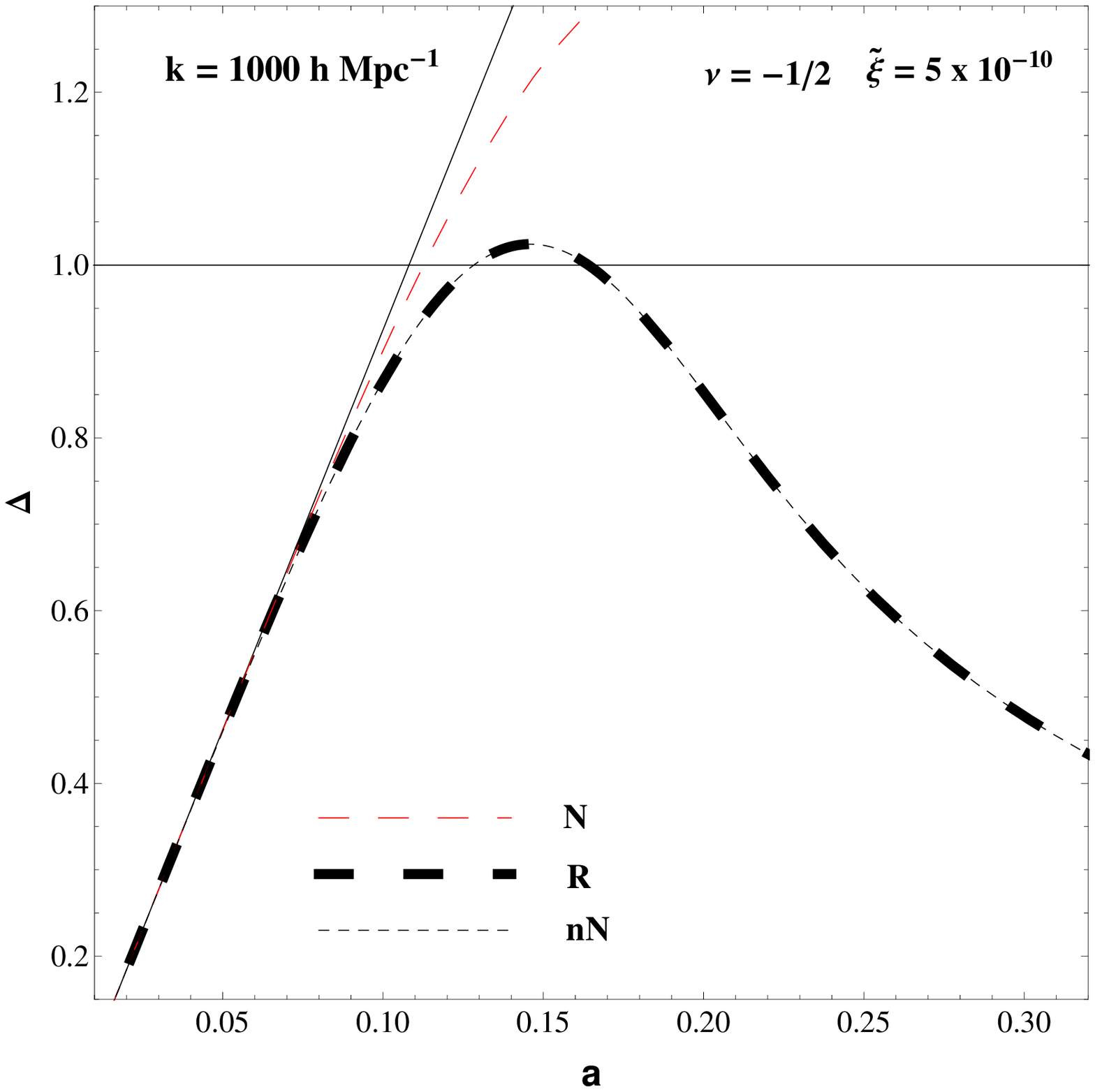}
\includegraphics[width=0.35\textwidth]{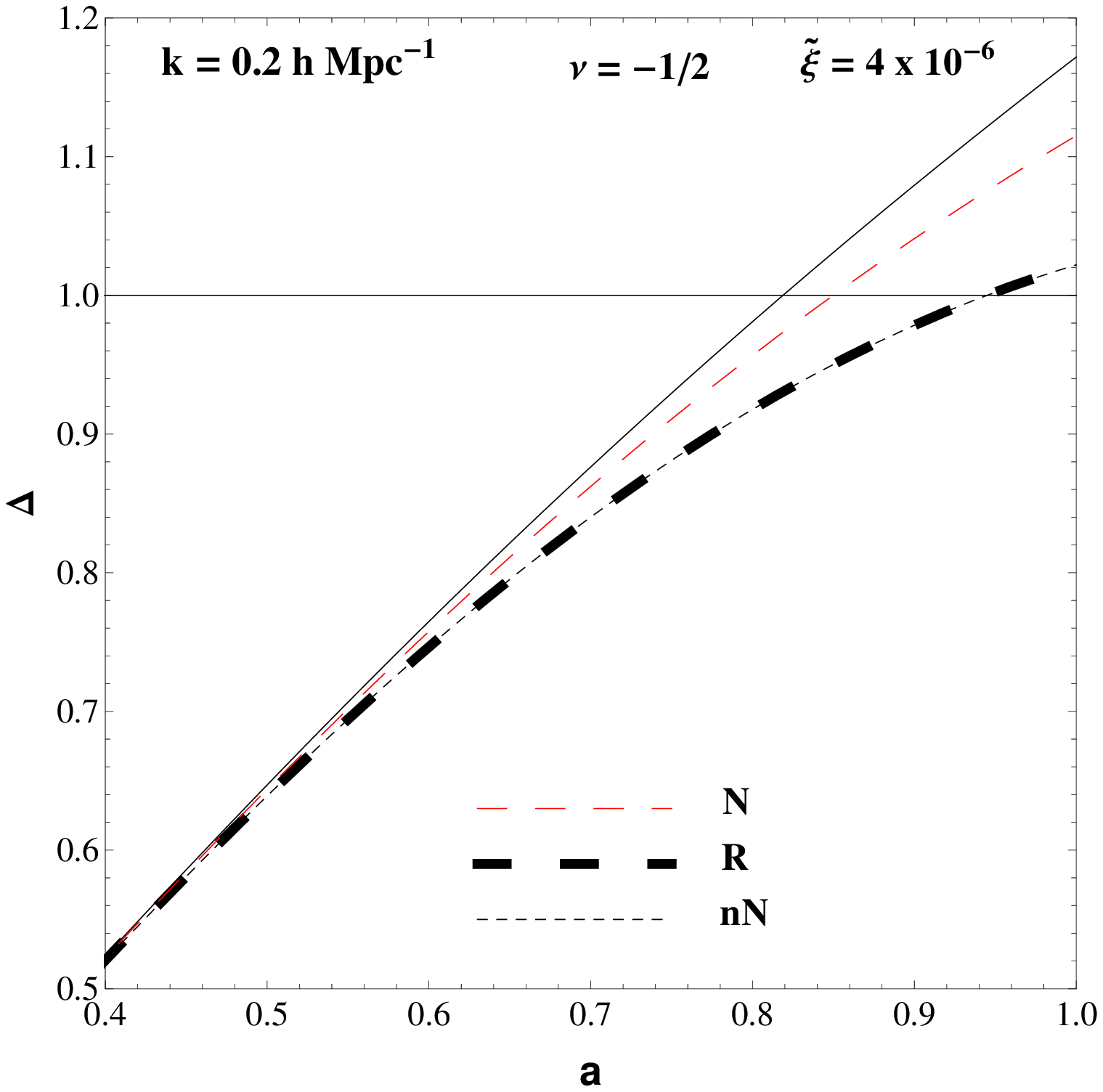}
\includegraphics[width=0.35\textwidth]{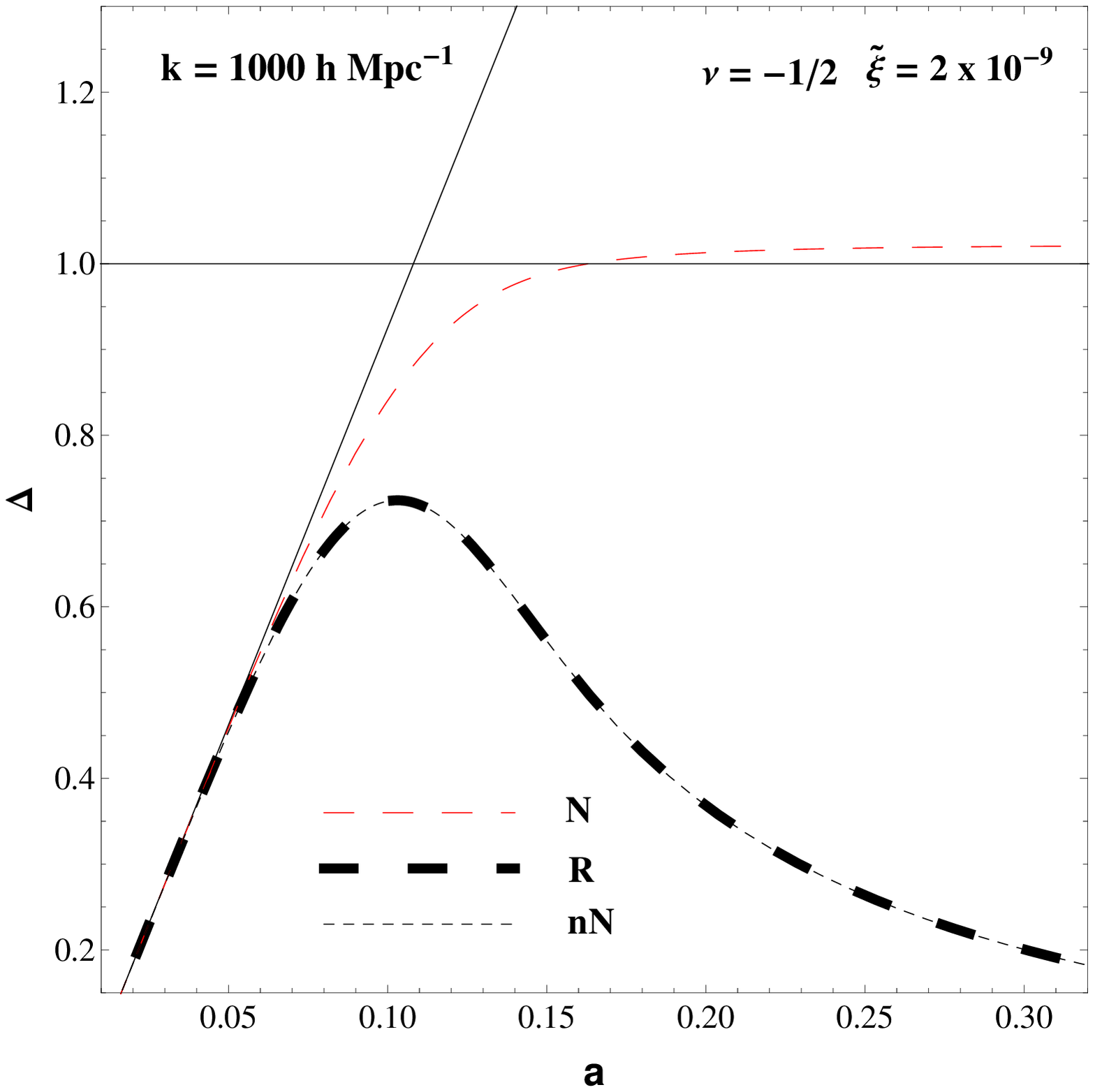}
\includegraphics[width=0.35\textwidth]{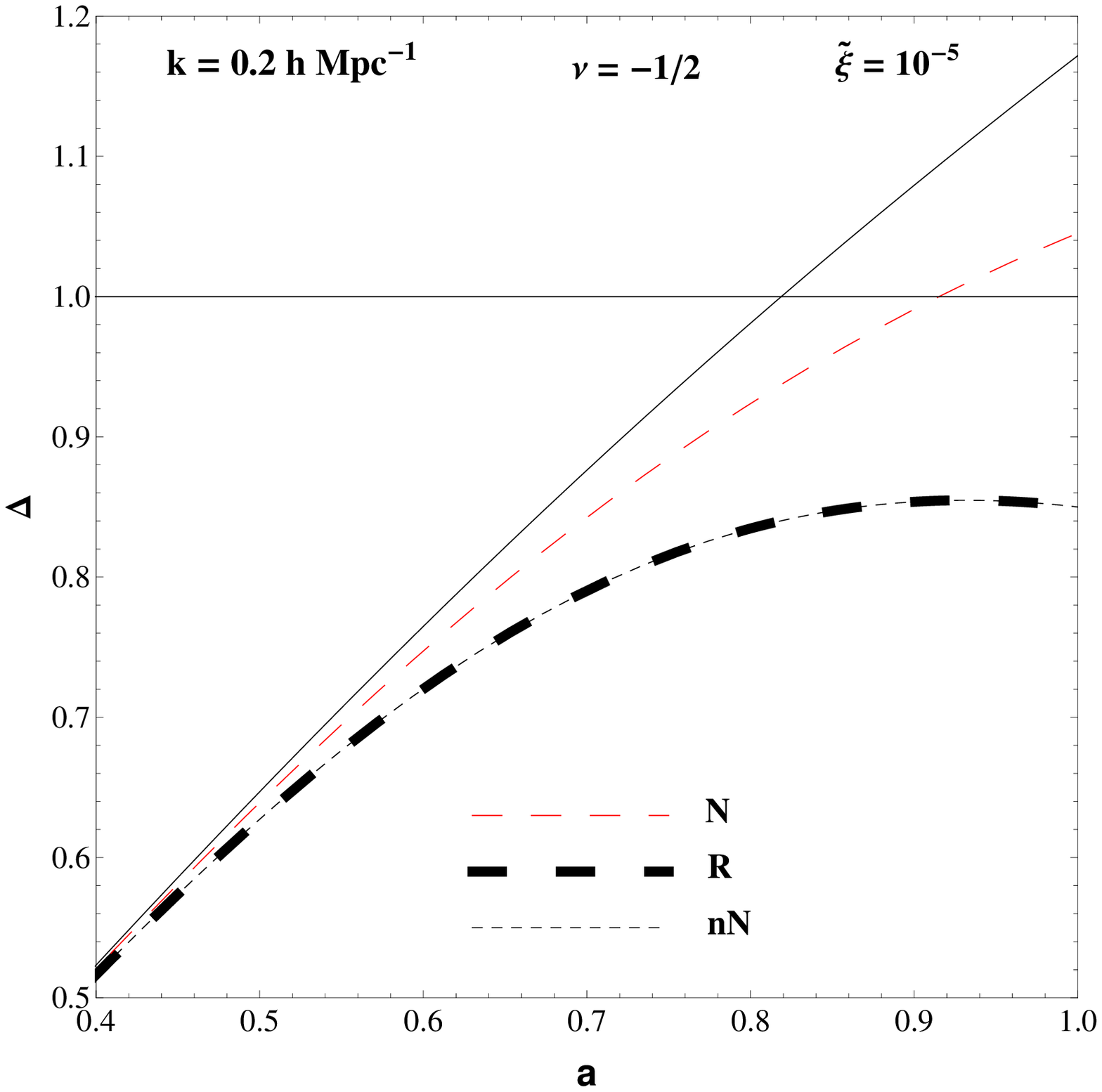}
\caption{As Fig.~1 but for $\nu=-1/2$ and different values for $\tilde{\xi}$. The neo-Newtonian and relativistic approaches agree with each other, but not with the Newtonian approach.}
\end{center}
\label{fig2}
\end{figure}

\begin{figure}
\begin{center}
\includegraphics[width=0.35\textwidth]{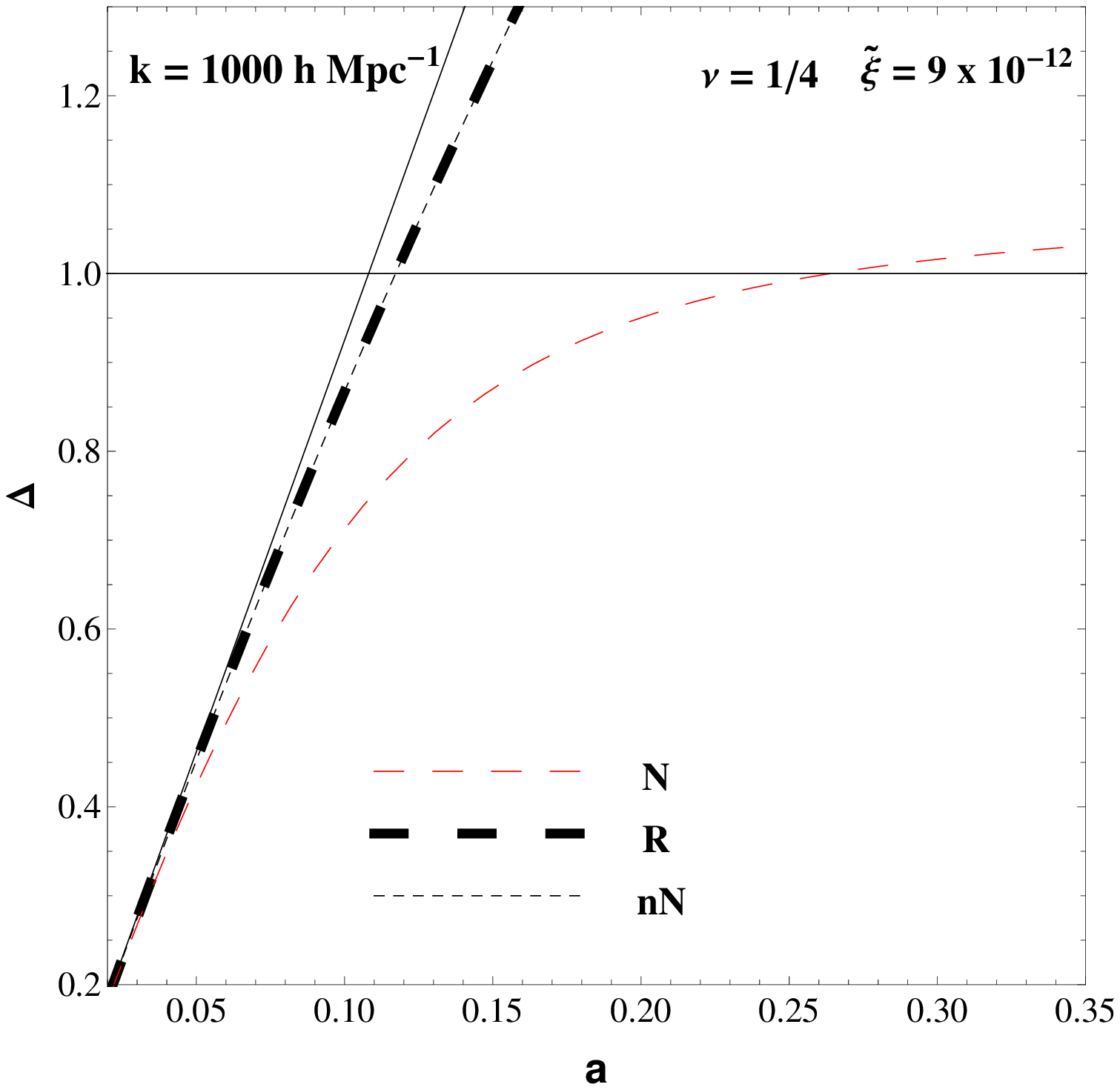}
\includegraphics[width=0.35\textwidth]{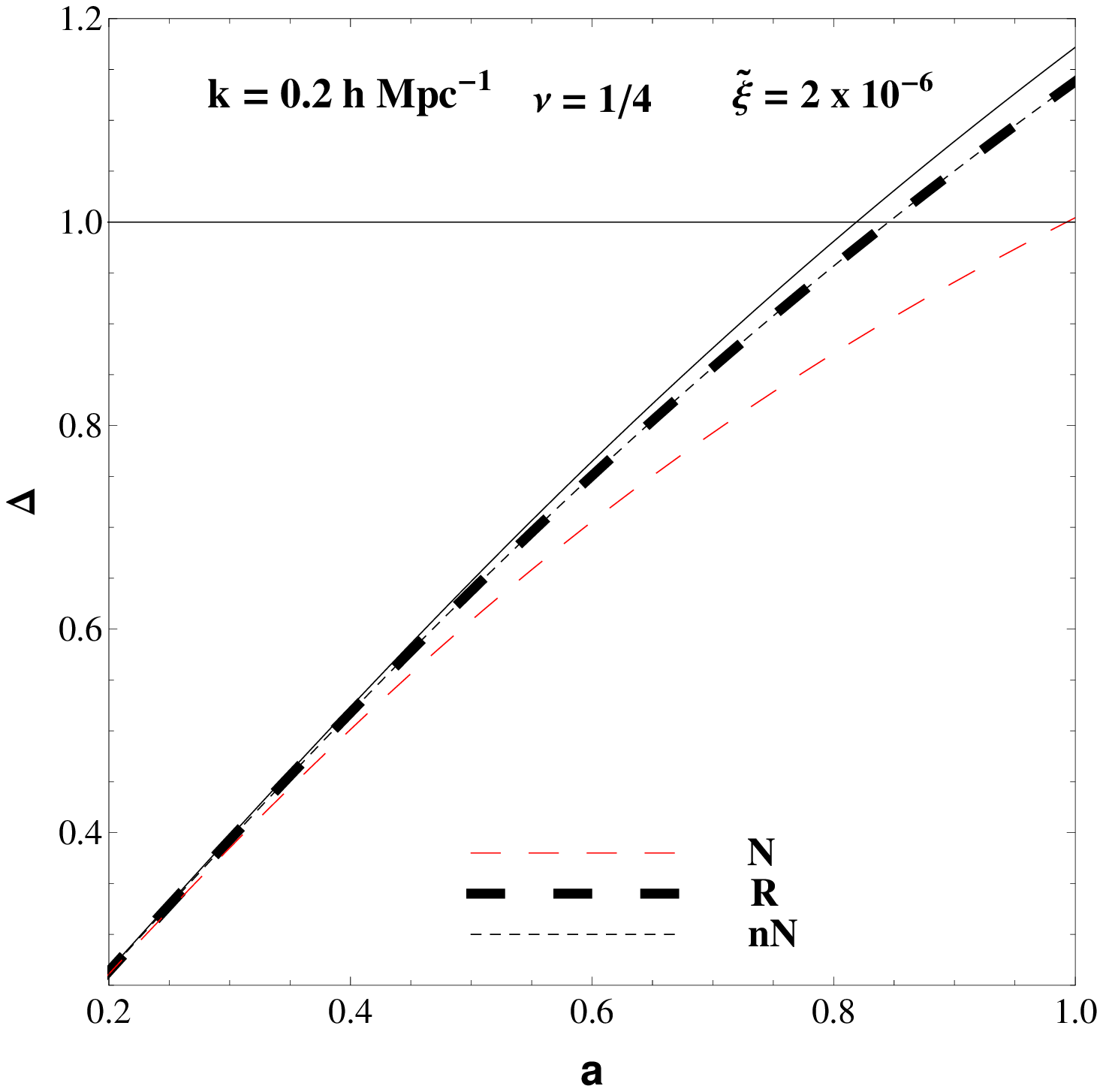}
\includegraphics[width=0.35\textwidth]{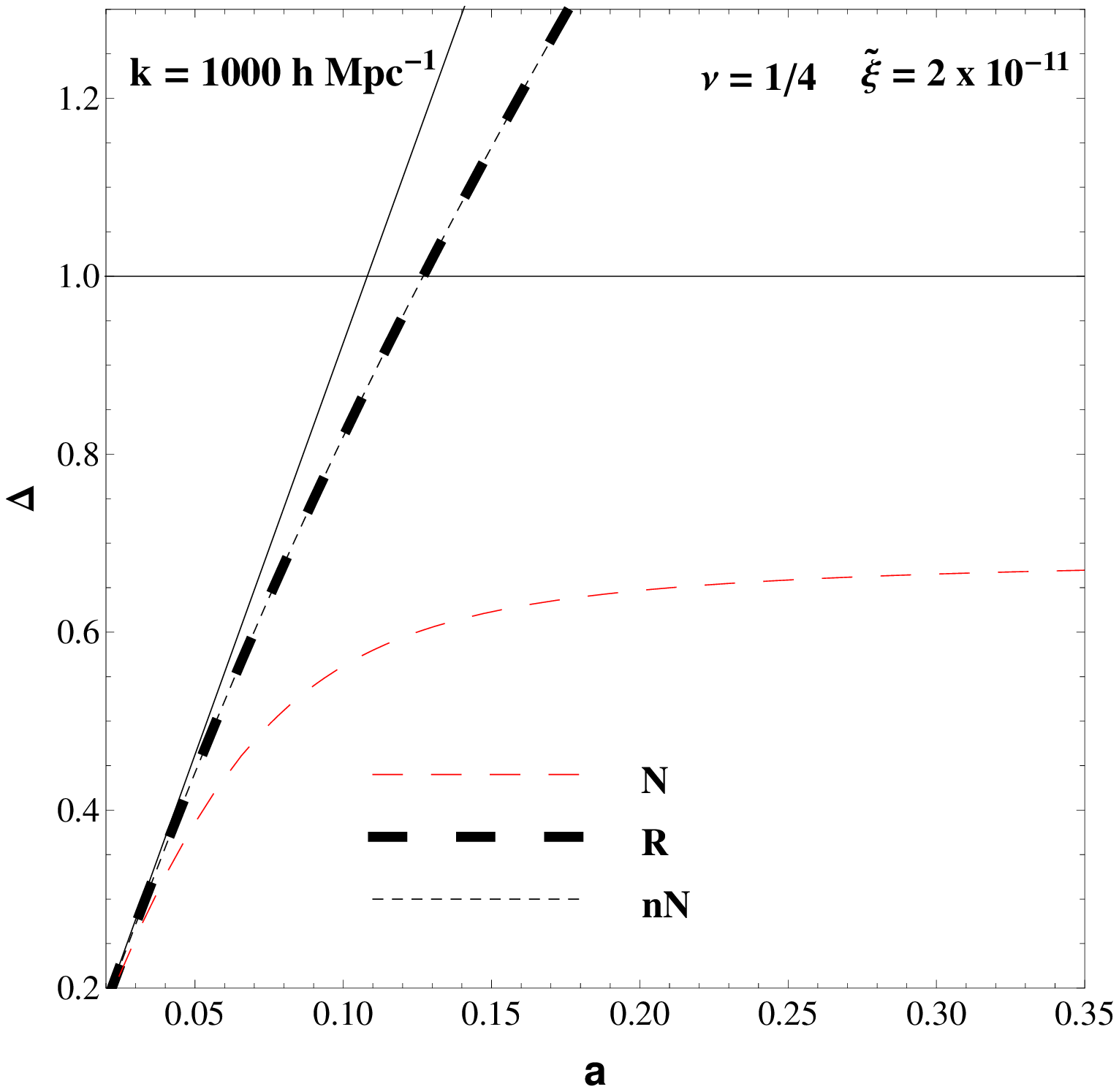}
\includegraphics[width=0.35\textwidth]{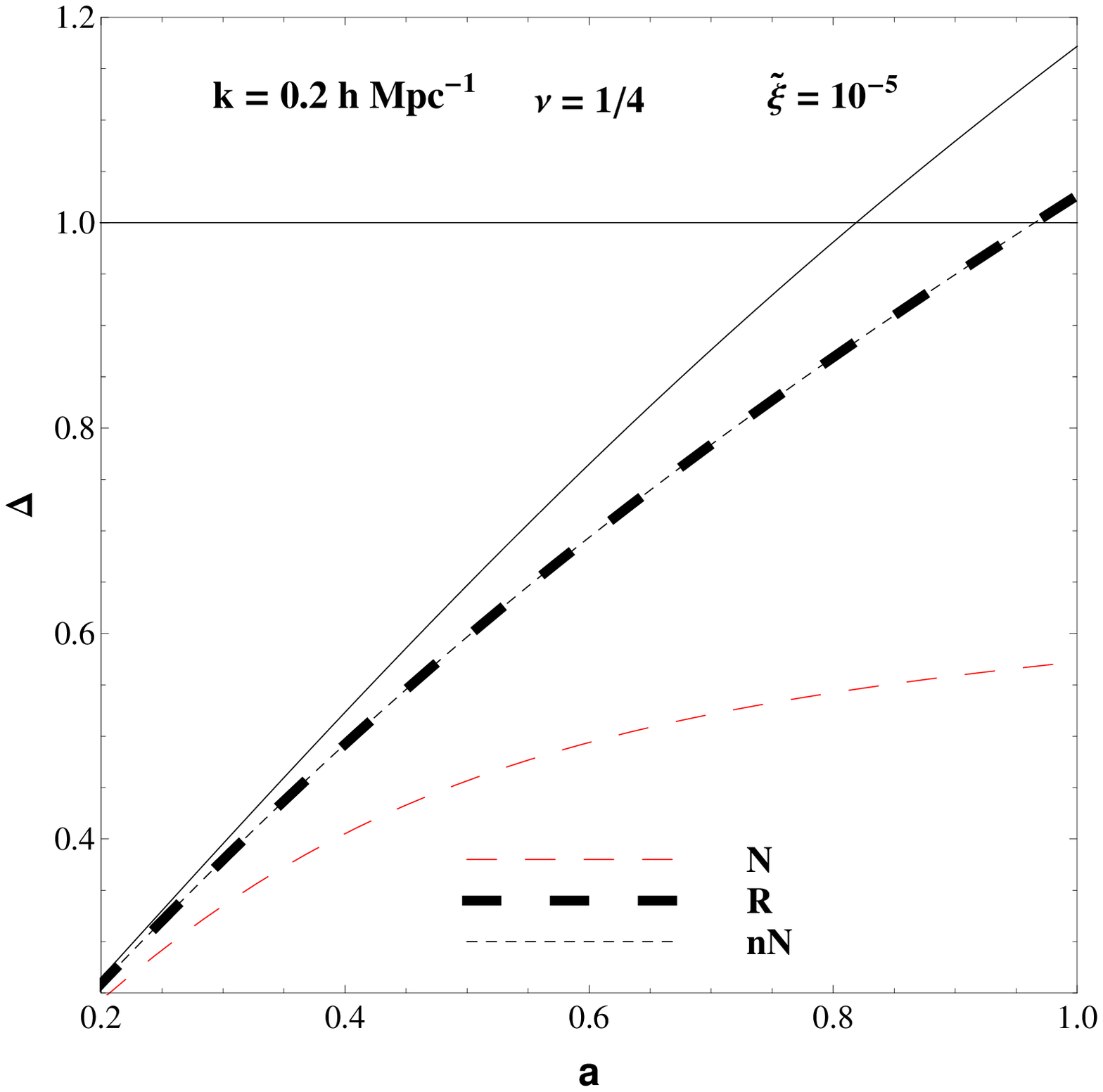}
\caption{As Fig.~2 but for $\nu=1/4$ and different values for $\tilde{\xi}$.}
\end{center}
\label{fig3}
\end{figure}

\begin{figure}
\begin{center}
\includegraphics[width=0.35\textwidth]{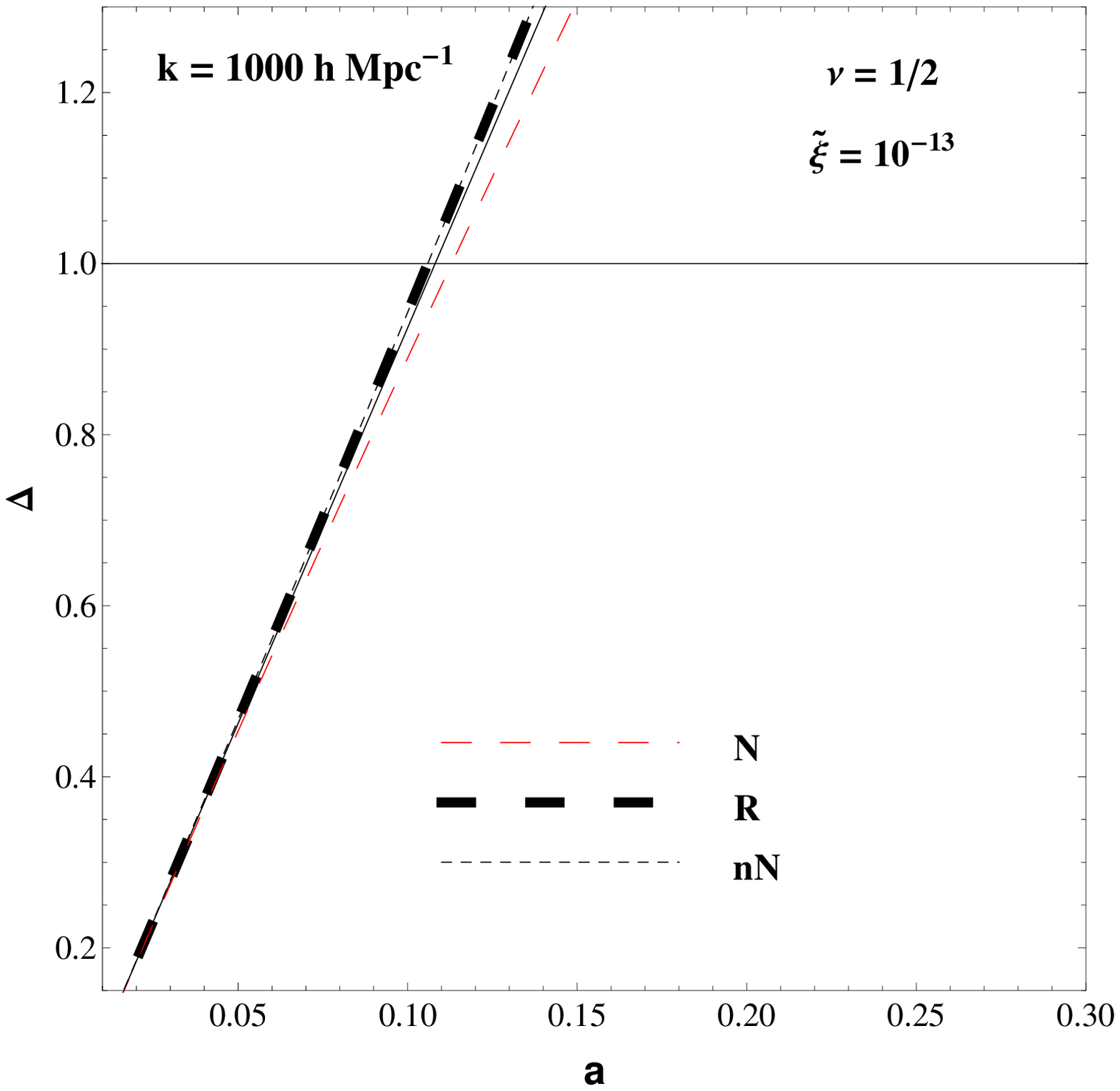}
\includegraphics[width=0.35\textwidth]{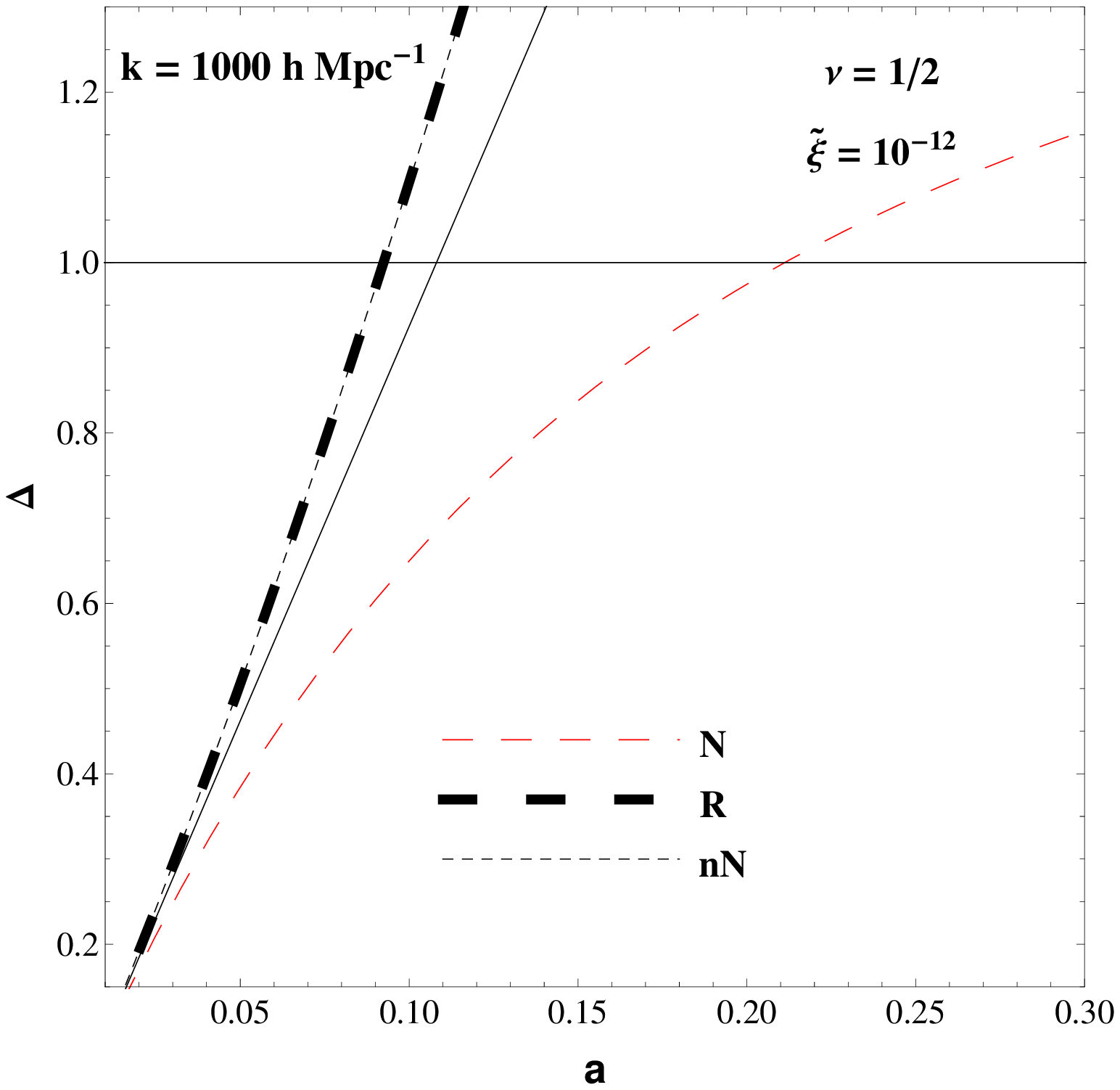}
\caption{Growth of sub-horizon density perturbations for the dwarf galaxy scale $k=1000 \, h \, {\rm Mpc}^{-1}$ assuming $\nu=1/2$. For this (unphysical) case bulk viscosity enhances the growth of structure.}
\end{center}
\label{fig4}
\end{figure}

\section{Conclusions}
\label{discussion}

Newtonian cosmology is a very useful tool to study both the linear and the non-linear evolution of density inhomogeneities in the universe. Cosmological N-body
and fluid simulations are based on Newtonian equations since i) the scales studied are well inside the Hubble horizon and ii) pressure effects are negligible. Both requirements are satisfied for the study of the growth of cold dark matter
halos at the galaxy cluster scale and below. In this work we address how the evolution of the dark matter density inhomogeneities is affected in the presence of viscous pressure.

Following \cite{dominikVelten2012}, we assigned a non-equilibrium pressure to dark matter in the
context of the $\Lambda$CDM model. We call this fluid the viscous cold dark matter ({\rm v}CDM). Here,
dark matter has a nonvanishing negative bulk viscous pressure, but it is not responsible for driving
the accelerated expansion of the universe. We thus have a $\Lambda$vCDM model. Since Newtonian
theory does not incorporate the effects of the pressure into the dynamics, the neo-Newtonian treatment
seems to be an appropriate approach to study the non-linear Newtonian evolution of structure
in this model. In order to validate this claim we compared the evolution of sub-horizon linear perturbations
in the presence of bulk pressure for the neo-Newtonian approximation and general relativity.

We derived the equations of motion for linear scalar perturbations for the {\rm v}CDM fluid
on sub-horizon scales for the Newtonian, neo-Newtonian and relativistic descriptions. We then focused on
the formation of dark matter halos for the dwarf galaxy and galaxy clusters scales.

The standard CDM growth is proportional to the scale factor $\Delta \propto a$ during the matter dominated
epoch. At late times and small scales, dark energy causes only a small attenuation of the growth of
structure. For the ${\rm v}$CDM scenario, we observe a strong growth suppression even for tiny values of the viscosity parameter $\tilde{\xi}$, consistent with the findings in \cite{dominikVelten2012}.

The key result of this work is that the dissipative, small-scale relativistic dynamics is fully described by the
neo-Newtonian theory for all values of the parameters $\nu$ and $\xi_0$.
Only in the exceptional case of a constant coefficient of bulk viscosity ($\nu=0$) all three approaches agree
and the usual Newtonian method applies. However, the range of validity of our results is limited only to very small values of the dark matter equation of state parameter. Indeed, this is the most interesting case because it is the situation where structures form.
Thus, apart from this case, Newtonian perturbation theory cannot be used to provide the correct growth of viscous CDM structures.
Consequently, a neo-Newtonian approach must be used when issues like the cuspy-core problem
or the missing satellite problem are addressed in numerical fluid simulations with modified CDM properties.

\section*{Acknowledgements}

This work was supported by CNPq (Brazil). HV and DJS acknowledge support from the DFG (Germany) within the Research Training Group 1620 ``Models of Gravity''.
JCF and WZ are also supported in part by FAPES (Brazil).

\end{document}